\documentclass[conference]{IEEEtran}
\IEEEoverridecommandlockouts
\usepackage{cite}
\usepackage{amsmath,amssymb,amsfonts}
\usepackage{algorithmic}
\usepackage{graphicx}
\usepackage{caption}
\usepackage{subcaption}
\usepackage{textcomp}
\usepackage[dvipsnames]{xcolor}
\usepackage[hidelinks]{hyperref}
\usepackage{url}
\usepackage{optidef}
\usepackage{tikz}
\usepackage{soul}
\def\BibTeX{{\rm B\kern-.05em{\sc i\kern-.025em b}\kern-.08em
    T\kern-.1667em\lower.7ex\hbox{E}\kern-.125emX}}

\newcommand{\xs}[1]{{\color{black} {#1}}}

\newcommand{\name}[1]{{MemorAI}}

\newcommand*\circled[1]{\tikz[baseline=(char.base)]{
            \node[shape=circle,draw,inner sep=2pt] (char) {#1};}}

\begin{document}

\renewcommand{\vec}[1]{\mathbf{#1}}

\title{MemorAI: Energy-Efficient Last-Level Cache Memory Optimization for Virtualized RANs\\
{\normalsize As accepted in the 2024 IEEE International Conference on Machine Learning for Communication and Networking (ICMLCN)}\vspace{-3mm}} 

\vspace{-11mm}

\author{\IEEEauthorblockN{Ethan Sanchez Hidalgo\IEEEauthorrefmark{2},
J. Xavier Salvat Lozano\IEEEauthorrefmark{1}, Jose A. Ayala-Romero\IEEEauthorrefmark{1},\\ Andres Garcia-Saavedra\IEEEauthorrefmark{1}, Xi Li \IEEEauthorrefmark{1} and
Xavier Costa-Perez\IEEEauthorrefmark{1}\IEEEauthorrefmark{2}\IEEEauthorrefmark{3}}
\IEEEauthorblockA{NEC Laboratories Europe\IEEEauthorrefmark{1},  i2CAT Foundation\IEEEauthorrefmark{2},  ICREA Foundation\IEEEauthorrefmark{3}\\ 
Email: ethan.sanchez@i2cat.net,
\{josep.xavier.salvat, jose.ayala, andres.garcia.saavedra, xi.li, xavier.costa\}@neclab.eu\\
}

\vspace{-10mm}

\thanks{The work was supported by the European Commission through Grant No.
SNS-JU-101097083 (BeGREEN). This work was also supported by the Spanish Ministry of Economic Affairs and Digital Transformation and the European Union – NextGeneration EU, in the framework of the Recovery Plan, Transformation and Resilience (PRTR) (Call UNICO I+D 5G 2021, ref. number TSI-063000-2021-3) Additionally, it has been supported by MINECO/NG EU (No. TSI-063000-2021-7) and the CERCA Programme.}}

\maketitle

\IEEEpubid{\begin{minipage}{\textwidth}\ \\[12pt]
\copyright2024 IEEE.  Personal use of this material is permitted.  Permission from IEEE must be obtained for all other uses, in any current or future media, including reprinting/republishing this material for advertising or promotional purposes, creating new collective works, for resale or redistribution to servers or lists, or reuse of any copyrighted component of this work in other works.
\end{minipage}} 

\IEEEpubidadjcol

\begin{abstract}
The virtualization of Radio Access Networks (vRAN) is well on its way to become a reality, driven by its advantages such as flexibility and cost-effectiveness. However, virtualization comes at a high price –- virtual Base Stations (vBSs) sharing the same computing platform incur a significant computing overhead due to in extremis consumption of shared cache memory resources. Consequently, vRAN suffers from increased energy consumption, which fuels the already high operational costs in 5G networks. This paper investigates cache memory allocation mechanisms' effectiveness in reducing total energy consumption. Using an experimental vRAN platform, we profile the energy consumption and CPU utilization of vBS as a function of the network state (e.g., traffic demand, modulation scheme).
Then, we address the high dimensionality of the problem by decomposing it per vBS, which is possible thanks to the Last-Level Cache (LLC) isolation implemented in our system. 
Based on this, we train a vBS digital twin, which allows us to train offline a classifier, avoiding the performance degradation of the system during training.
Our results show that our approach performs very closely to an offline optimal oracle, outperforming standard approaches used in today's deployments.

\end{abstract}


 \section{Introduction}\label{secc:intro}
 
RAN virtualization is a crucial technology for reducing the Total Cost of Ownership (TCO) of 5G RAN infrastructure~\cite{masoudi2020cost, murti2020optimal, samsung_whitepaper}. Virtualized RANs (vRANs) are expected to import the advantages of network function virtualization (NFV), such as exploiting general-purpose computing platforms and shortening deployment cycles. Collaborative efforts like the carrier-led O-RAN alliance have stimulated the market and the research community to develop innovative solutions that incorporate the adaptability and cost-effectiveness of NFV right into the edge of mobile networks \cite{oran-cloud-scenarios}. Nonetheless, while shared computing platforms offer enhanced flexibility and cost-effectiveness, they also introduce challenges for 5G base stations, as they compromise the predictability offered by dedicated platforms~\cite{tootoonchian2018resq, garcia2021nuberu, kumar2019picnic}. The term \emph{noisy neighbor problem} has been coined to refer to the issue when shared resources are consumed in extremis, meaning that another function restricts one virtualized function’s resources. This problem has motivated substantial research over the years~\cite{garcia2021nuberu, subramanian2015application, park2019copart, selfa2017application}.

\begin{figure}[t!]
 \centering
\minipage{0.45\columnwidth}
 \includegraphics[width=\columnwidth]{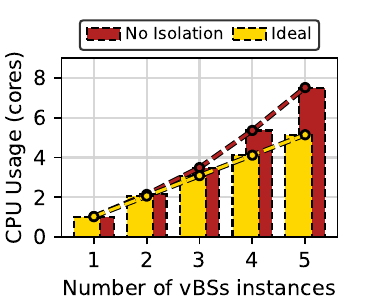}
\vspace{-5mm}
\caption{\small Aggregated per-core usage with \# of vBS instances in our PoC vRAN platform.}
 \label{fig:intro:vran}
\endminipage{}
\hfill
\minipage{0.45\columnwidth}
\includegraphics[width=\columnwidth]{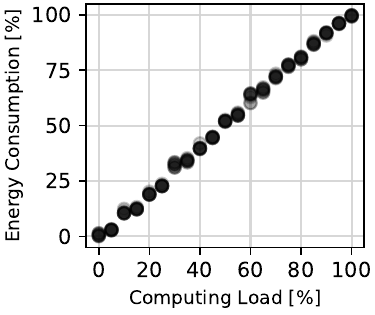}
\vspace{-5mm}
\caption{\small Energy consumption as a function of the computing load.}
 \label{fig:intro:energy}
\endminipage{}
\vspace{-6mm}
\end{figure}

Virtual RANs are not alien to the \emph{noisy neighbors problem} as shown in our previous work  \cite{salvat2023airic}. When deploying multiple virtual Base Stations (vBSs) instances in a vRAN system, the usage of computing resources increases. We confirm this by deploying multiple vBS instances using Docker containers into a pool of computing cores in a shared off-the-shelf server. In contrast to \cite{salvat2023airic}, in these experiments we configured the pool of computing cores to retain as much predictability as possible (see \S\ref{secc:exp_analysis} for more details). 

Fig.~\ref{fig:intro:vran} depicts the aggregated computing usage per core in our vRAN platform as a function of the number of deployed vBS under maximum traffic load in uplink (UL) and downlink (DL). The bars in yellow show the expected usage assuming perfect resource isolation in place. We compute these by linearly scaling up the CPU usage of a single vBS instance. The red bars show the actual CPU consumption, which unveils an exponentially-growing overhead induced by the aforementioned \emph{noisy neighbors problem}.

As we explained in~\cite{salvat2023airic}, the increased computing overhead has its roots in the lack of cache memory isolation. The computing overhead poses an issue of the increase of the total energy consumption of the vRAN platform. Fig.~\ref{fig:intro:energy} shows the relationship between the normalized energy consumption on top of the system's baseline (i.e. idle) consumption of our vRAN platform as a function of the total computing load. The computing load and the energy are linearly related~\cite{fan2007power, lefurgy2007server}. Therefore, it is key to minimize the computing usage of our vRAN platform to keep operational energy costs low. \\

In our previous work~\cite{salvat2023airic}, we developed a solution to dimension the computing capacity of a vRAN system, considering the increased computing usage due to the \emph{noisy neighbors problem}. However, this work does not consider minimizing the increased overhead but adapting to it. In this paper, we seek to reduce the computing overhead as much as possible. We begin studying how isolating the different cache memory levels influences the total system computing consumption. We found that vBSs traffic demand is virtually orthogonal to using cache resources, i.e., vBSs use as much cache memory as possible. However, the utility of cache memory is different for different traffic demands. The vBSs with higher demands and Signal-to-Noise-Ratio (SNR) can reduce their computing usage more when they have more cache memory available. Thus, developing a solution that strategically allocates the cache memory resources to the different vBS instances according to their demands is key to minimize energy consumption. Due to the complex relationship between computing and radio resources~\cite{ayala2019vrain}, we propose a novel approach that can effectively allocate the cache resources to minimize the total computing usage and consequently reduce the energy consumption of vRAN platforms.
\section{Background: memory in general-purpose computing platforms}\label{sec:background}

Cache memory~\cite{jacob2010memory} bridges the speed gap between RAM and the CPU itself and it is organized into various levels based on speed and size. The first two cache memory levels called L1 and L2 cache are the closest and fastest to the system, although its capacity is limited. Each physical core has its dedicated L1 and L2 cache. L1 is faster than L2 but lower in size. Lastly, the L3 cache, also known as the \emph{Last Level Cache (LLC)}, is the slowest CPU's cache and is shared among all computing cores\cite{patterson2003modern, jacob2010memory, drepper2007every}. Table~\ref{tab:cache_cycles} shows the access latency and the available size per core of different cache levels in the Intel Skylake architecture. 

\begin{table}[t!]
\footnotesize
\centering
\begin{tabular}{c|c|c|}
Memory type & Access latency\cite{patterson2003modern, jacob2010memory, drepper2007every} & Size per core \\\hline
L1 cache & 4-6 cycles & 64 KB \\
L2 cache & 14 cycles & 	256 KB\\
L3 (LLC) cache & 50-70 cycles & 2 MB\\
RAM & $\sim$ 120 - 600 cycles & 2-4 GB\\
\end{tabular}
\caption{\small Access and cache miss latency.}
\label{tab:cache_cycles}
\vspace{-6mm}
\end{table}

Cache memory is organized into several associative sets also known as cache ways. A cache way is a set of lines or blocks that can store data from a specific location in the main memory. Tools such as Intel CAT\footnote{\href{https://github.com/intel/intel-cmt-cat}{https://github.com/intel/intel-cmt-cat}} enables allocating LLC cache ways to specific cores or processes.

In a general-purpose computing platform, when a core executing a thread requires frequently used memory blocks, it loads them into the cache for quicker access. Consequently, if a thread references a memory block that is not present in the cache, the core initiates an interrupt known as a ``cache miss'' and searches for the data in a higher memory level. This incurs additional CPU cycles which increase the total execution time of a computing thread.

\section{Experimental analysis}\label{secc:exp_analysis}

\subsection{Cache memory isolation}

\begin{figure}[t!]
 \centering
 \includegraphics[width=0.9\columnwidth]{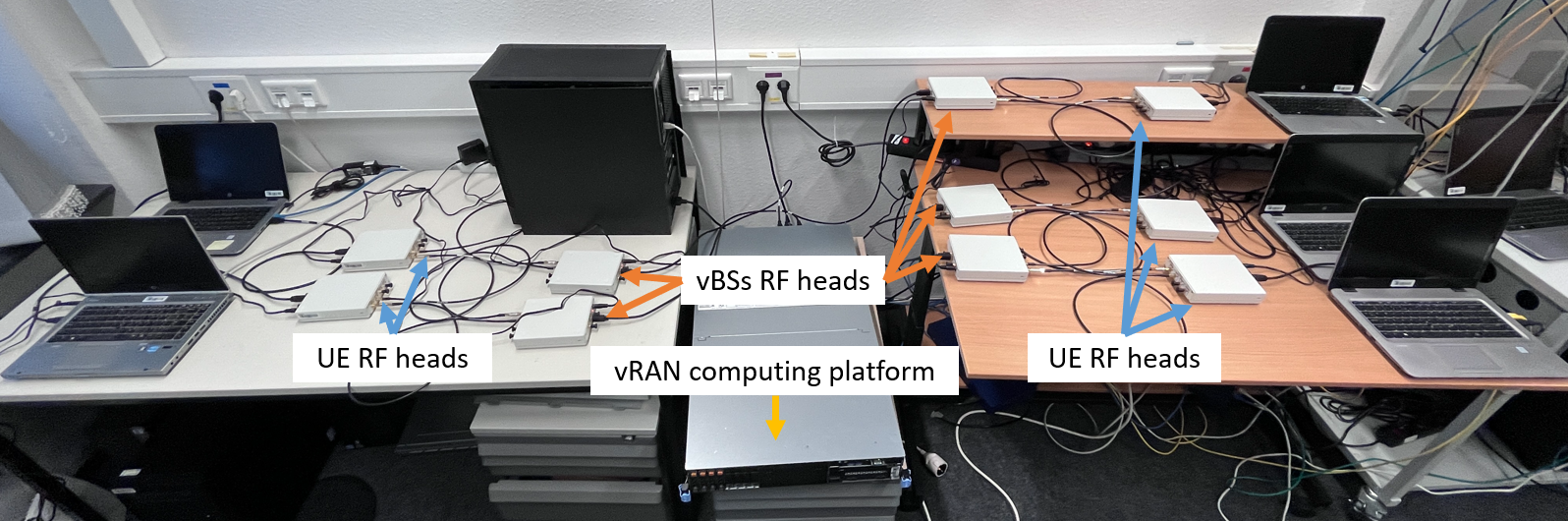}
 \vspace{-2mm}
 \caption{\small vRAN testbed.}
 \label{fig:diss:testbed}
\vspace{-6mm}
\end{figure}

To evaluate the impact of cache memory resources on the \emph{noisy neighbor problem} in vRANs’ energy consumption, we measure the computing usage and low-level cache metrics deploying a different number of vBS instances in different scenarios in our vRAN platform. Fig.~\ref{fig:diss:testbed} shows our experimental vRAN platform. Therein, we deploy the different vBS instances in an isolated pool of computing cores from an Intel Xenon E5-2650 v4 CPU @ 2.20GHz in a shared off-the-shelf server. Each vBS has its dedicated RF radio head connected to one UE, which is used to emulate the aggregated cell load. We configured the pool of computing cores to retain as much predictability as possible: ($i$) we isolated $12$ cores with $12$ dedicated Last Level Cache (LLC) ways, ($ii$) the system can only use C0/C1 C-states and we turned off hardware P-states and ($iii$) we deactivated Hyper-threading. We then initiate bidirectional data flows, both uplink (UL) and downlink (DL), with maximum load and good wireless channel conditions between each vBS instance and different user equipment (UE). We consider the following scenarios:

\begin{itemize}
    \item \emph{Ideal}. 
    We compute the CPU usage scaling linearly the usage of a single vBS instance assuming that the cache memory size also scales linearly.
    \item \emph{No isolation}. We deploy an increasing quantity of vBS instances without any cache memory isolation mechanism.
    \item \emph{Pinning.} L1 and L2 cache levels are dedicated per core. We pin different vBS instances to distinct cores to assess the impact on L1 and L2 cache isolation. To facilitate comparisons, the number of cores assigned to each vBS is given by: 
    \begin{equation}
        \text{cores per vBS} = \bigg\lfloor \frac{\text{total cores}}{\text{deployed vBSs}} \bigg\rfloor
    \end{equation}
    where the total number of cores equals $12$ and the number of deployed vBS increases from $1$ to $5$.
    Note that when 5 vBSs are deployed, each vBS is assigned to 2 cores and two free cores are left.
    \item \emph{Pinning + LLC isolation}. We perform the L3 cache allocation in the same manner as the CPU pinning. 
    We allocate the total number of cache ways equally for every single vBS, i.e. the number of cache ways per vBS is given by:
     \begin{equation}
        \text{cache ways per vBS} = \bigg\lfloor \frac{\text{total cache ways}}{\text{deployed vBSs}} \bigg\rfloor
    \end{equation}
In our platform, there are a total of 12 cache ways. We use Intel CAT to allocate the LLC cache ways to the corresponding computing cores.

\end{itemize}

To measure the computing usage we use the \texttt{/proc} filesystem to read and store periodically the computing time for the different threads of a vBS on the computing cores allocated. On the other hand, we used the tool \texttt{perf} to measure the number of instructions per cycle (IPC) and the number of cache misses per 1k instructions (MPKI) by one vBS.

\begin{figure}[t!]
 \centering
 \includegraphics[width=0.40\columnwidth]{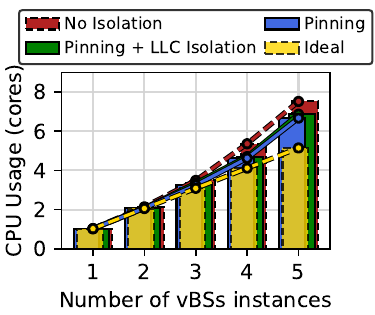}
 \vspace{-2mm}
 \caption{\small Comparison of the aggregated per-core usage with \# of vBS instances showing the ``No isolation'', the ``Pinning'' and the ``Pinning + LLC isolation'' scenarios.}
 \label{fig:diss:cum_cpu_usage_comparison}
\vspace{-5mm}
\end{figure}

\begin{figure}[t!]
\vspace{-0mm}
\minipage{0.47\columnwidth}
\centering
 \includegraphics[width=0.82\columnwidth]{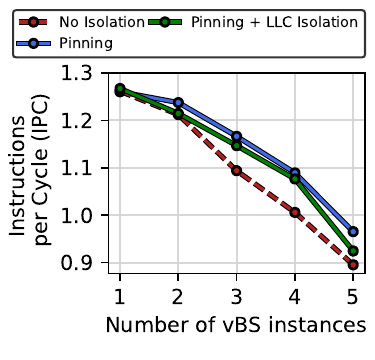}
 \vspace{-4mm}
 \caption{\small Instructions per cycle (IPC) with \# of vBSs.}
 \label{fig:diss:ipc_comparison}
 \vspace{-7mm}
\endminipage
\hfill
\minipage{0.47\columnwidth}
 \includegraphics[width=0.82\columnwidth]{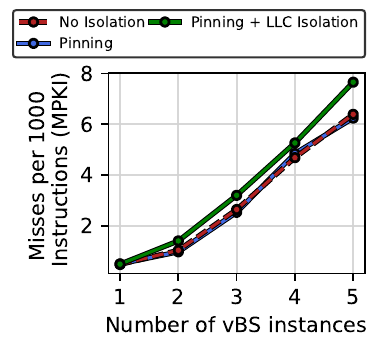}
 \vspace{-4mm}
 \caption{\small Misses per 1000 instruction (MPKI) with \# of vBSs.} 
 \label{fig:diss:mpki_comparison}
 \vspace{-7mm}
\endminipage
\end{figure}

Fig.~\ref{fig:diss:cum_cpu_usage_comparison} shows the measured computing usage in the different scenarios.
We observe that, for the no isolation configuration, the computing usage increases by approximately $50\%$ compared to the ideal case, increasing the energy consumption. This also impacts the IPC and MPKI, as shown in Fig.~\ref{fig:diss:ipc_comparison}-\ref{fig:diss:mpki_comparison}, respectively. We observe that the IPC decreases and the cache misses increase as more instances are deployed. There is a sixfold increase in the cache misses when transitioning from $1$ to $5$ vBSs. This increase also impacts the number of cycles required to execute the same number of instructions, decreasing the IPC.

The \emph{Pinning} configuration shows a lower computing usage than the \emph{No isolation} but still higher than the ideal case. The computing usage decrease is significant considering that L1 cache ways are approximately $100$ times lower in size than L3 cache ways and L2 cache ways are $10$ times lower in size than L3 cache ways. In Fig.~\ref{fig:diss:ipc_comparison} we can observe a higher IPC across any number of deployed vBSs compared to the \emph{No isolation} scenario. However, Fig.~\ref{fig:diss:mpki_comparison} shows the same number of cache misses for all the cases. This might seem shocking at first glance, but as we have previously detailed L1 and L2 cache way size is very low compared to L3. 

Finally, to study the effect of the L3 cache isolation. In Fig.~\ref{fig:diss:cum_cpu_usage_comparison}, the \emph{Pinning + LLC isolation} configuration does not improve the computing usage. Specifically, the computing usage is the same for all the cases except for the case with $5$ vBS, in which it is slightly higher. The reason behind this behavior is that we only allocate $10$ out of $12$ L3 cache ways to all vBSs for the case of $5$ vBSs, while in the \emph{Pinning} configuration, all vBSs have all the L3 cache memory available, using on average $2.4$ L3 cache ways.

\subsection{LLC occupancy and utility}

\begin{figure}[t!]
\centering
    \begin{subfigure}[t!]{0.43\columnwidth}
        \includegraphics[width=0.85\columnwidth]{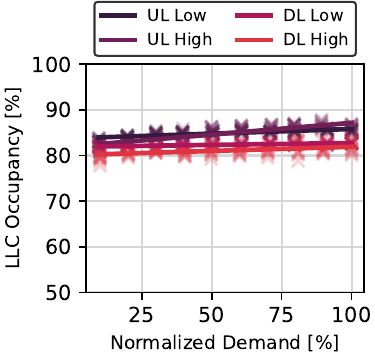}
        \vspace{-2mm}
        \caption{\small $12$ LLC cache ways.}
        \label{fig:diss:llc_12}
    \end{subfigure}
    \begin{subfigure}[t!]{0.43\columnwidth}
        \includegraphics[width=0.85\columnwidth]{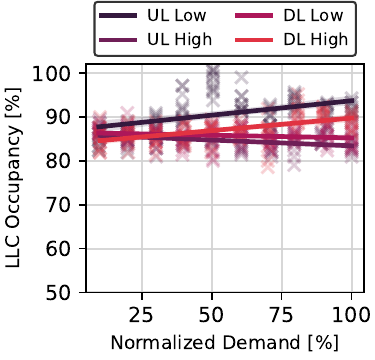}
        \vspace{-2mm}
        \caption{\small $2$ LLC cache ways.}
        \label{fig:diss:llc_2}
    \end{subfigure}
\caption{LLC occupancy in \% as a function of the total demand for different SNR cases in UL and DL.}
\label{fig:diss:llc}
\vspace{-5mm}
\end{figure}

We now study how the allocation of LLC cache ways impacts the computing usage of a vBS instance. First, we measure the percentage of LLC cache memory used of the total LLC memory allocated to a vBS as a function of the traffic demand. Fig.~\ref{fig:diss:llc} shows the LLC occupancy with $12$ and $2$ LLC cache ways for different SNR values and traffic demands in uplink and downlink. The \emph{high} series depict the LLC occupancy with a high SNR environment while the \emph{low} series depict the LLC occupancy with a low SNR environment. We can see that the total LLC occupancy is above $80\%$ for uplink and downlink. Also, the LLC occupancy is almost orthogonal to the demand of the vBS, yielding an increase of a $2-6\%$ when the demand goes from $10\%$ to $100\%$.

\begin{figure}[t!]
 \centering
 \includegraphics[width=0.75\columnwidth]{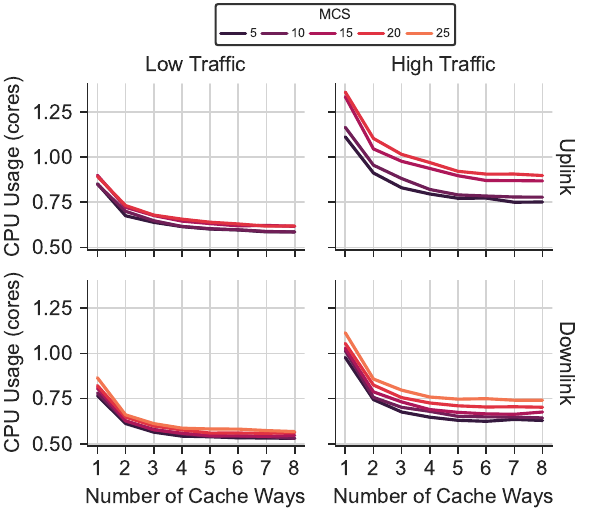}
 \vspace{-2mm}
 \caption{\small Computing usage as a function of the LLC allocated cache ways for different SNR and in UL and DL.}
 \label{fig:intro:cpu_mcs_traffic_comparison}
\vspace{-6mm}
\end{figure}

Finally, Fig.~\ref{fig:intro:cpu_mcs_traffic_comparison} depicts the computing usage of a vBS as a function of the L3 cache ways when we deploy it using $3$ cores. Fig.~\ref{fig:intro:cpu_mcs_traffic_comparison} depicts the computing usage for the different Modulation Coding Schemes (MCSs) with a low traffic demand (i.e. 20\% of the total demand) and high traffic demand (i.e. 100\% of the total demand) for uplink and downlink. We observe that there are significant differences between the achievable computing usage reduction. For high traffic both in uplink and downlink, we can achieve a more significant computing usage reduction. On the contrary, a vBS that processes a low traffic demand can attain lower gains in terms of computing usage. We conclude that LLC resources have different \emph{utility} depending on the vBS context.

This contrasts with Fig.~\ref{fig:diss:llc} which shows that the vBS makes full use of the cache memory regardless of the demand. Thus, if there is no LLC cache allocation mechanism, all vBSs deployed will be using the same amount of LLC cache memory on average. It is key to strategically distribute the LLC cache ways among the vBSs deployed boosting its utility to minimize the computing usage and therefore the energy consumption depending on the traffic demands.

\subsection{The problem}

\begin{figure}[t!]
\centering
    \begin{subfigure}[t!]{0.45\columnwidth}
        \includegraphics[width=0.75\columnwidth]{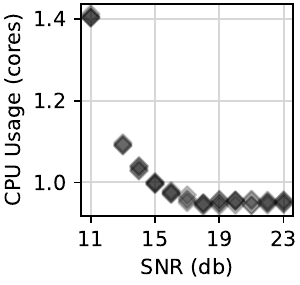}
        \vspace{-2mm}
        \caption{\small Computing usage.}
        \label{fig:model:cpu}
    \end{subfigure}
    \begin{subfigure}[t!]{0.45\columnwidth}
        \includegraphics[width=0.75\columnwidth]{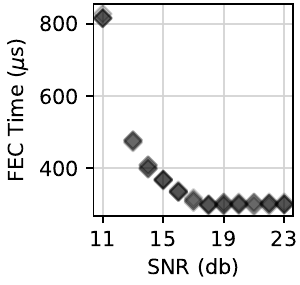}
        \vspace{-2mm}
        \caption{\small Decoding time.}
        \label{fig:model:dec_time}
    \end{subfigure}
\vspace{-1mm}
\caption{Computing usage and decoding time of a vBS with max. UL and DL load with mild MCS over different SNR conditions.}
\vspace{-6mm}
\end{figure}

Quantifying the LLC cache memory utility is a challenging task. The LLC cache utility depends on the computing demands of the vBS which are influenced by various factors, including traffic demand in both the downlink (DL) and uplink (UL), the signal-to-noise ratio (SNR) of each wireless link, and the specific Modulation Coding Scheme (MCS) utilized for communication. All these elements interact in a complex manner~\cite{ayala2019vrain},~\cite{garcia2021nuberu}. Fig.~\ref{fig:model:cpu} depicts the relative mean core usage of the vBS, and shows that, given a MCS, lower SNR regimes demand a higher amount of computing resources. The underlying reason is the iterative nature of the forward-error-correction (FEC) algorithms -- signals received with lower SNR require a higher number of FEC iterations to decode the transported codeword successfully. This is confirmed by Fig.~\ref{fig:model:dec_time}, which shows the amount of time taken by the decoder to finish its task for every transport block.

\section{Problem Formulation}

We consider a vRAN platform comprising $M_{\textnormal{cores}}$ computing cores and $N_{\textnormal{LLC}}$ LLC cache ways. We consider that $N_{\textnormal{vBS}}$ vBS instances are deployed in the platform. Every vBS $i$ in the vRAN platform has both a UL and DL traffic demand denoted by $d^{\textnormal{UL}}_{i}$ and $d^{\textnormal{DL}}_{i}$, respectively; a SNR $s_{i}$; and an MCS in UL and DL denoted by $m^{\textnormal{UL}}_{i}$ and $m^{\textnormal{DL}}_{i}$, respectively, that the radio scheduler selects depending on $s_{i}$. The vBS $i$ has a set of isolated cores $P_i$ such that $\vert P_i \vert > 0$ and a number of dedicated LLC cache ways $n^{LLC}_{i}$, where $n^{LLC}_{i} \geq 1$ $\forall i$. We need to satisfy that $\sum_i \vert P_{i} \vert \leq M_{cores}$ and $\sum_i n^{LLC}_{i} = N_{LLC}$. We define $c_{i}$ as the computing use of vBS $i$.

We define $\vec{x}_{i} := (d^{\textnormal{UL}}_{i}, d^{\textnormal{DL}}_{i}, s_{i}, m^{\textnormal{UL}}_{i}, m^{\textnormal{DL}}_{i})$ as the context of the vBS $i$. Moreover, we define $f_i$ as the function that maps $\vec{x}_{i}$ and $n^{LLC}_{i}$ to the computing usage $c_{i}$. 

Finally, we define vector $\vec{\mathcal{X}}$ which concatenates the context vectors from all vBS as $\vec{\mathcal{X}} := (\vec{x}_{1}, \vec{x}_{2}, \dots \vec{x}_{N_{\textnormal{vBS}}})$. Also, we define $\mathcal{P} := (P_1, \dots, P_{N_{\textnormal{vBS}}})$ and $\mathcal{N} := (n^{LLC}_{1}, \dots, 
 n^{LLC}_{N_{\textnormal{vBS}}})$ as the vectors with the core set allocation and the LLC cache ways allocation on a vRAN platform. Similarly, we define the function $f$ which maps $(\mathcal{X}, \mathcal{N})$ to the total computing usage $C^{\textnormal{vRAN}} \in [0, M_{cores}]$ of the vRAN platform. We define the problem of optimizing the computing set and the LLC allocation as: 

\vspace{-5mm}

\begin{align}\label{eq:optProb}
&\! \min_{\mathcal{N}}        &\qquad& f(\mathcal{X}, \mathcal{N})\\
&\text{subject to} &      & \sum_i n^{LLC}_{i} = N_{LLC}. \nonumber
\end{align}

\vspace{-2mm}

As explained in \S\ref{secc:intro}, the computing usage and the energy consumption are proportional. Thus, minimizing the computing usage function $f$ also minimizes the total energy consumption. Note that, in our problem, the assignment of CPU cores to vBS $\mathcal{P}$ is already given. We select a fixed value of $\mathcal{P}$ based on our experimental insights and previous works on this topic~\cite{salvat2023airic, ayala2019vrain} (see Sec. \ref{sec:perf_eval} for more details). Note that, in the problem \eqref{eq:optProb}, the optimal LLC cache allocation depends on the context $\mathcal{X}$, whose dimensionality increases depending on the number of vBSs. Moreover, the optimal action is also dependent on the number of active vBS, which may change over time. To avoid the exploration burden of learning algorithms (e.g., reinforcement learning) that can lead to suboptimal configurations, we decompose the problem and use a digital twin system.

\section{\name{}}

\begin{figure}[t!]
 \centering
 \includegraphics[width=0.75\columnwidth]{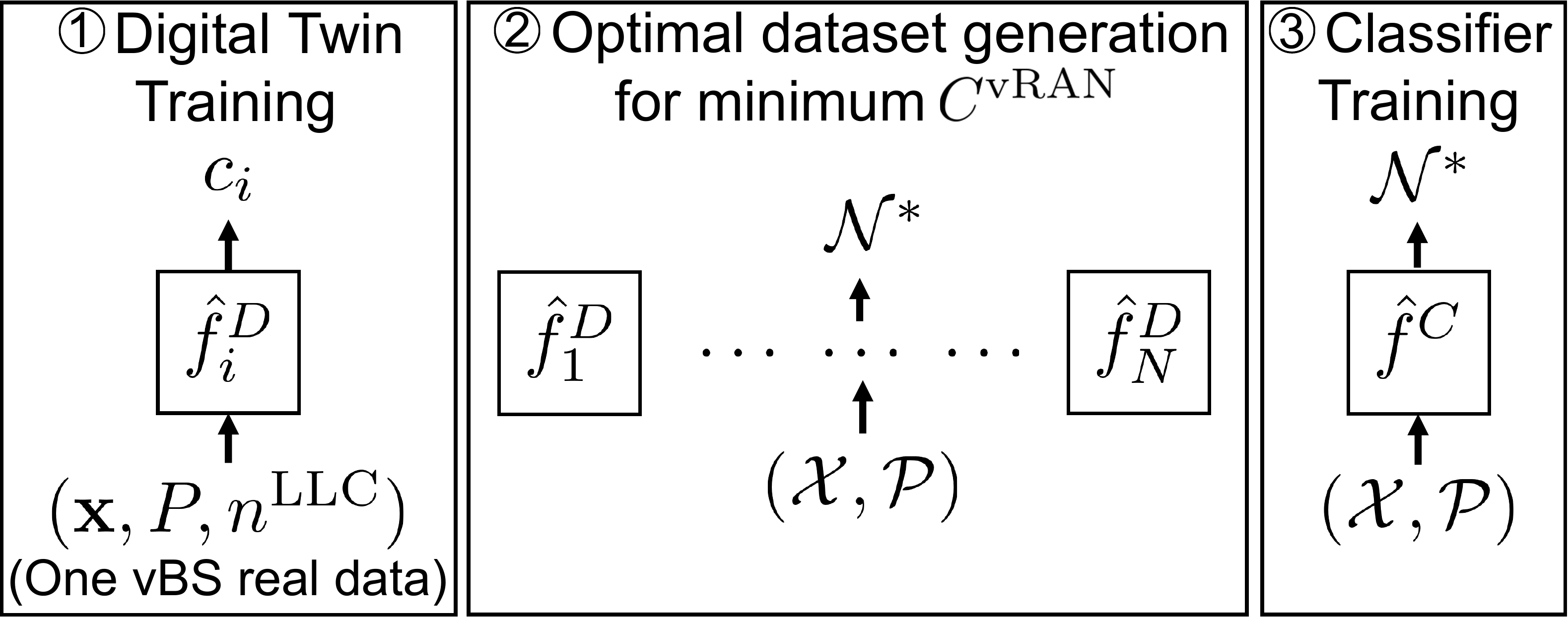}
 \vspace{-1mm}
 \caption{\small MemorAI optimization framework.}
 \label{fig:model:memorai}
\vspace{-6mm}
\end{figure}

To solve~(\ref{eq:optProb}), we propose an optimization framework which we call \textit{\name{}}. This framework considers discrete decision intervals denoted by $t \in \{1, 2, \dots, T\}$.  At the beginning of each decision interval, our solution receives $\vec{\mathcal{X}}^{(t)}$, and decides the optimal LLC allocation $\mathcal{N}^{*(t)}$ across all vBSs, to minimize the energy consumption. \textit{\name{}} is composed of a \xs{set of} Digital Twin\xs{s} (DT\xs{s}) and a Neural Network (NN) classifier. Fig.~\ref{fig:model:memorai} shows our optimization framework.

\subsection{Digital Twin}

Thanks to full vBS isolation via pinning it to dedicated cores and LLC cache ways allocation, we make the observation that $C^{\textnormal{vRAN}} = \sum_{i}c_{i}$ as there are no joint effects between vBSs. This observation implies that $f(\mathcal{X}, \mathcal{N}) = \sum_{i}f_i(\vec{x}_{i}, n^{LLC}_{i})$. As there is no interaction among vBS in terms of computing usage, we can create DTs of independent vBS. Thus, we can mirror their behavior in a safe and controlled environment for testing and learning. Each DT can model the particularities of each vBS (e.g., different implementations protocol stacks) and we can emulate the complex interactions of the full vRAN system. We create a DT using operational data from one vBSs. Using each vBS’s DTs, we can lower the time cost to generate a dataset for a number of vBS as we aggregate the results of each DT. Note that without the DTs, the amount of data needed to create a training dataset increases exponentially with the number of vBS (curse of dimensionality).

To create the Digital Twin of one vBS we used our vRAN platform to generate a training dataset with a fixed computing core set $\vert P \vert$. Note that we select a set of cores such that vBS can correctly operate. This makes $f_i$ a continuous function. Each dataset sample contains a $4$-tuple with $(\vec{x}, P, n^{\textnormal{LLC}}, c)$. Using this dataset, we built up a DT using a Neural Network (NN) which approximates $c$ using $(\vec{x}, P, n^{\textnormal{LLC}})$ i.e. it approximates $f_i$ minimizing the Minimum Squared Error (MSE). We denote the DT function as $\hat{f}_i^D$. Fig.~\ref{fig:model:memorai} \circled{1}, shows the DT training step.

\subsection{NN Classifier}
The DT allows us to evaluate the CPU usage of different configurations very accurately without having to use the real system. Therefore, we can perform an exhaustive search to find the optimal LLC allocation $\mathcal{N}^*$ for a set of contexts $(\mathcal{X}, \mathcal{P})$. \xs{Note that the size of the set of possible LLC cache allocations is $|\mathcal{N}| = \binom{N_{LLC} - 1}{N_{vBS} - 1}$.} Fig.~\ref{fig:model:memorai} \circled{2}, shows how using the different DTs we generate the previous data set.

Using this data set, we train a fully connected NN to predict $\mathcal{N}^*$ for a given context $(\mathcal{X}, \mathcal{P})$. Specifically, we solve a multi-class classification problem using the cross-entropy loss function. We denote the classifier function as $\hat{f}^C$. Note that our solution is very flexible to changes in the system (e.g., upgrades in the implementation of the software stack, deployment of new vBSs, etc.). In those cases, after having the digital twin modeling, we can easily retrain the NN classifier offline without degrading the performance of the system. Finally, Fig.~\ref{fig:model:memorai} \circled{3}, shows the classifier training step of our optimization framework.

\section{Performance Evaluation}

\begin{figure}[t!]
 \centering
\minipage{0.47\columnwidth}
 \includegraphics[width=0.95\columnwidth]{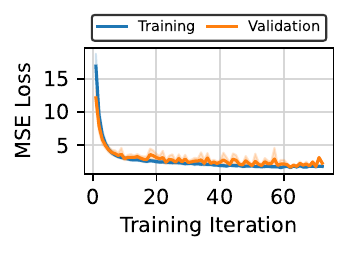}
\vspace{-3mm}
\caption{\small Digital Twin MSE Loss.}
 \label{fig:perf_eval:mse_digital_twin}
\endminipage{}
\hfill
\minipage{0.47\columnwidth}
\includegraphics[width=0.95\columnwidth]{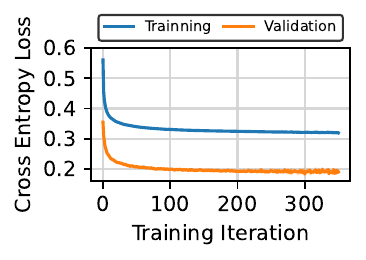}
\vspace{-3mm}
\caption{\small NN Classifier Cross Entropy Loss.}
 \label{fig:perf_eval:learning_curve_classifier}
\endminipage{}
\vspace{-6mm}
\end{figure}

\label{sec:perf_eval}

In this section, we evaluate the performance and potential savings of our approach.
We carry out the evaluation for a $N_{vBS} = 5$ vBS deployment. We used PyTorch\footnote{\href{https://pytorch.org/}{https://pytorch.org/}} to implement the Digital Twin and the Classifier.

\subsection{Training Evaluation}

\subsubsection{Digital Twin}

We implemented the DT of one vBS using a NN with three hidden layers of sizes $\{256, 128, 64\}$ respectively with a ReLU activation function. Moreover, we stop the training iterations using an early stopping mechanism to prevent overfitting \cite{prechelt2002early}. The early stopping mechanism stops the training after the value of the loss function in a validation data set has not improved during the last $N$ training iterations. $N$ is usually referred to as patience.

Fig.~\ref{fig:perf_eval:mse_digital_twin} shows the MSE loss value on the training and validation data sets of the DT as a function of the training iterations. We selected a patience of $N = 10$ and trained the model during $70$ iterations.

\subsubsection{NN Classifier}
On the other hand, we implemented the Classifier as a NN with 4 hidden fully connected layers of sizes $\{512, 384, 384, 512\}$ respectively. Each layer also used a ReLU activation function and we introduced a $0.2$ dropout probability during training. We also use the early stopping mechanism with $N = 50$ of patience.

Fig.~\ref{fig:perf_eval:learning_curve_classifier} shows the cross entropy loss on the training and validation data set as a function of the number of training iterations. We train the model until iteration $300$ (due to the early stopping mechanism). The training loss is higher than the validation loss due to the dropout layers. Also, the classifier achieves a $92.1\%$ accuracy on our testing data set.

\subsection{Performance Benchmark}

\begin{figure}[t!]
\centering
    \begin{subfigure}[t!]{0.45\columnwidth}
        \includegraphics[width=1.\columnwidth]{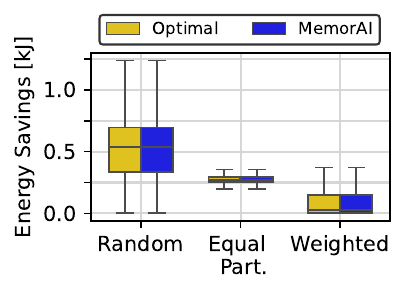}
        \vspace{-6mm}
        \caption{\small 12 Cache Ways.}
    \label{fig:perf_eval:savings_random_context}
    \end{subfigure}
    \begin{subfigure}[t!]{0.45\columnwidth}
        \includegraphics[width=1.01\columnwidth]{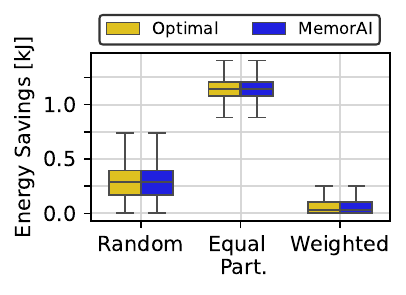}
        \vspace{-6mm}
        \caption{\small 8 Cache Ways.}
        \label{fig:perf_eval:savings_high_traffic}
    \end{subfigure}
\vspace{-2mm}
\caption{\small Energy savings compared to different benchmarks and different number of cache ways for a 15 min decision interval.}
\vspace{-6mm}
\end{figure}

We design \textit{\name{}} to operate in the Non-Real Time RIC of the O-RAN architecture as an rApp~\cite{nonrtric}. Based on this, we selected a time granularity of 15 minutes for our evaluation \cite{marquez2018should}. We generated a data set with random context data and compared \textit{\name{}} against the following approaches:
\begin{itemize}
    \item \textbf{Random}: We select the allocation of cache ways for each vBS randomly; 
    \item \textbf{Equal partition}: All the vBSs get allocated the same number of cache ways. Extra cache ways are left unallocated;
    \item \textbf{Weighted}: Each vBS gets allocated a number of cache ways as proportionally to its total demand as:
    
    $$ n^{LLC}_{i} = \frac{d^{\textnormal{UL}}_{i} + d^{\textnormal{DL}}_{i}}{\sum_j d^{\textnormal{UL}}_{j} + d^{\textnormal{DL}}_{j} } \cdot N_{\text{LLC}}$$
        
\end{itemize}

Fig.~\ref{fig:perf_eval:savings_random_context} shows the energy savings in kilojoules (kJ) of our solution and the optimal strategy with respect to the different benchmarks in a decision interval of 15 minutes. Our solution outperforms the three benchmarks in terms of energy while producing almost the same results as the optimal solution. Thus, \textit{\name{}} understands better the utility of LLC cache partitioning than benchmark strategies.

Our solution yields higher savings up to $1$ kJ compared to the random strategy, up to $0.35$ kJ compared to the equal partitioning, and up to $0.36$ kJ compared to the weighted strategy. 
Note that, although the weighted strategy shows a lower power consumption, it does not scale the LLC cache properly when there is an imbalance between UL and DL demands. In these cases, our approach achieves the highest gains with respect to this strategy.

Finally, Fig.~\ref{fig:perf_eval:savings_high_traffic} shows the attained gains when the system has only $8$ cache ways available. In that case, the random approach performs better because the number of configurations is lower and therefore the probability of selecting the optimal configuration is higher.
The savings compared to the equal partition approach are higher as the benchmark can only allocate one cache way per vBS in this scenario. Compared to the weighted strategy, we still observe higher savings.

\section{Related Work}

The \emph{noisy neighbor problem} has been studied 
for different types of virtualized resources. The works in~\cite{khalid2018iron, kumar2019picnic} propose 
different solutions to effectively isolate the networking stack. The authors in~\cite{khalid2018iron} show that containers' computing time processing packets via interrupts is not correctly accounted. 
Therefore, the authors develop a solution to called Iron which effectively charges the computing time used for processing packets to each container. Authors in~\cite{kumar2019picnic} 
develop PicNIC a predictable virtualized NIC abstraction that effectively provides predictable performance to cloud providers. 

The works in~\cite{subramanian2015application, selfa2017application, park2019copart} develop different solutions to partition the memory resources to effectively isolate different tenants using the same infrastructure. The work in~\cite{subramanian2015application} 
develops a solution to estimate the slowdown of an application due to the interference from other applications and propose different strategies
to share the memory resources.
In~\cite{selfa2017application} authors 
develop a clustering solution to fairly partition the LLC cache ways for different applications using Intel CAT. 
Moreover, in~\cite{park2019copart} the authors develop a solution to fairly allocate LLC cache and memory bandwidth for workload consolidation.

Finally,~\cite{salvat2023airic, garcia2021nuberu, foukas2021concordia} tackle the \emph{noisy neighbor problem} 
in the context of vRANs. In our previous work in~\cite{salvat2023airic} we showed that vRANs experience an increased computing consumption due to interference in the cache memory that can lead users to lose connectivity. We designed AIRIC, an AI controller that dimensions the computing capacity of a vRAN system considering the computing overhead. 
The authors in~\cite{garcia2021nuberu} enhance the physical layer pipeline of operations of a vBS 
so that it is suitable to run in non-deterministic computing platforms. 
Finally, in~\cite{foukas2021concordia} authors develop a solution to increase the CPU utilization in vRAN platform opportunistically co-locating non-5G workloads while ensuring correct operation. 

\section{Conclusions}

Cache memory is a key resource for vRANs to reduce energy consumption. Non-isolated access to cache memory resources increases energy consumption, making less attractive the advantages of virtualization. In our work, we have studied how the different mechanisms for cache memory isolation decrease the energy consumption of a vRAN platform. Then, we proposed \textit{\name{}} which strategically allocates LLC resources to minimize energy consumption. \textit{\name{}} comprises a digital twin and a neural network classifier, providing a very efficient and flexible solution. \textit{\name{}} achieves almost optimal performance and can attain significant energy savings when compared with other strategies.
\bibliographystyle{IEEEtran}
\bibliography{references}

\end{document}